\documentclass[twocolumn,tightenlines,floats,floatfix,prc,nofootinbib,preprintnumbers]{revtex4}

\def\bea{\begin{eqnarray}}
\def\eea{\end{eqnarray}}

\def\st#1{{\kern-4pt} \not\!#1}

\def\sp{\kern +3pt}
\def\sm{\kern -3pt}

\usepackage[usenames]{color}

\def\be{\begin{equation}}
\def\ee{\end{equation}}
\def\ba{\begin{eqnarray}}
\def\ea{\end{eqnarray}}

\usepackage{graphics}
\usepackage{graphicx}
\usepackage{epsf} 
\usepackage{amsmath}
\usepackage{amssymb}
\usepackage{slashed}
\usepackage{bm}

\begin{document}

\phantom{0}
\vspace{-0.2in}
\hspace{5.5in}

\preprint{ {}}

\vspace{-1in}

\title
{\bf Parametrizations of the $\gamma^\ast N \to \Delta(1232)$ 
quadrupole form factors \\ 
and Siegert's theorem}
\author{G.~Ramalho
\vspace{-0.1in} }

\affiliation{International Institute of Physics, 
Federal University of Rio Grande do Norte, 
Campus Universit\'ario - Lagoa Nova, CP.~1613,
Natal-RN, 59078-970, Brazil}

\vspace{0.2in}
\date{\today}

\phantom{0}

\begin{abstract}
The large $N_c$ limit provides relations 
that can be used to calculate the 
$\gamma^\ast N \to \Delta(1232)$ quadrupole form factors 
at low and intermediate $Q^2$
under the assumption of the pion cloud dominance.
There are two limitations in those parametrizations.
First, the parametrization of the Coulomb 
quadrupole form factor underestimate the low $Q^2$ data.
Second, when extrapolated for the timelike region, 
the form factors violate  
Siegert's theorem by terms of the order $1/N_c^2$.
We propose here corrections to the   
parametrization of the electric quadrupole form factor, 
which violate Siegert's theorem only 
by terms of the order $1/N_c^4$.
Combining the improved large $N_c$ pion cloud 
parametrizations with the valence quark contributions 
based on a covariant quark model 
for the quadrupole transition form factors,
we obtain an extrapolation to the timelike region 
consistent with Siegert's theorem, and 
accomplish also a very good description of the data.
\end{abstract}

\vspace*{0.9in}  
\maketitle

\section{Introduction}

The  $\gamma^\ast N \to \Delta(1232)$ transition 
is characterized by the dominant 
dipole form factor ($G_M$) and two 
sub-leading quadrupole form factors:
the electric ($G_E$) and the Coulomb ($G_C$) 
form factors~\cite{Jones73,Pascalutsa07b,NDeltaD,NSTAR}.
The nonzero results for the quadrupole form factors 
are a consequence of the deviation of the $\Delta(1232)$
from a spherical shape~\cite{Pascalutsa07b,Deformation,Quadrupole}.

Calculations based on the limit of a large number of colors ($N_c$)
and $SU(6)$ quark models with symmetry breaking suggest that,
in the low $Q^2$ region the $\gamma^\ast N \to \Delta(1232)$ 
quadrupole form factors are dominated by 
pion cloud effects~\cite{Pascalutsa07b,Pascalutsa07a,Buchmann97a,Grabmayr01,Buchmann04,Buchmann02,Buchmann09a}.
Estimates based exclusively on valence quarks 
underestimate the data by about  
an order of magnitude~\cite{Capstick90,JDiaz07,Stave08,NDeltaD,NSTAR}.
Although small comparative to 
the leading order pion cloud contributions,
the valence quark contributions can
nevertheless help to improve the description 
of the data~\cite{LatticeD,JDiaz07,Kamalov99}.
The large $N_c$ parametrizations 
of the $\gamma^\ast N \to \Delta(1232)$
electric and Coulomb quadrupole form factors 
have, however, a problem:  they are in conflict with  
Siegert's theorem~\cite{Buchmann98,Drechsel2007,SiegertD,Siegert}.

In the form factors representation, 
Siegert's theorem is expressed by 
the identity at the pseudo-threshold:  
$G_E =\frac{M_\Delta -M}{2 M_\Delta}  G_C$~\cite{SiegertD,Siegert,Jones73}
($M$ and $M_\Delta$ are respectively the nucleon and the $\Delta$ masses).
The pseudo-threshold is the limit where 
the magnitude of the photon three-momentum, $|{\bf q}|$, vanishes,
and $Q^2= Q_{pt}^2= -(M_\Delta - M)^2$.
A test for the  validity of  
Siegert's theorem is the value of
\ba
{\cal R}_{pt}= G_E(Q_{pt}^2) - \kappa G_C( Q_{pt}^2),
\label{eqR}
\ea
where 
\ba
\kappa = \frac{M_\Delta -M}{2 M_\Delta}.
\ea
When ${\cal R}_{pt}=0$, Siegert's theorem is verified.
When ${\cal R}_{pt} \ne 0$, the form factors are 
inconsistent with Siegert's theorem.

The combination of $SU(6)$ quark models with 
two-body exchange currents and the large $N_c$ limit 
provides a connection between the neutron charge 
distribution and the $\gamma^\ast N \to \Delta(1232)$ 
quadrupole form factors~\cite{Buchmann09a,Buchmann04}.
In an exact $SU(6)$ quark model, the neutron 
electric form factor vanishes  
and the electric and Coulomb quadrupole 
moments of the  $\gamma^\ast N \to \Delta(1232)$ transition 
are both zero.
When the  $SU(6)$ symmetry is broken 
we can relate the quadrupole moments 
with the neutron square charge radius 
$r_n^2$~\cite{Pascalutsa07a,Buchmann97a,Grabmayr01,Buchmann04,Buchmann09a,Buchmann02}.
We can then conclude 
that the $SU(6)$ symmetry breaking induces
an asymmetric distribution of charge in the nucleon, 
which generates nonzero results for the   
neutron electric form factor $G_{En}$,  
and for the  $\gamma^\ast N \to \Delta(1232)$ 
quadrupole transition form factors, 
$G_E$ and $G_C$~\cite{Buchmann97a,Grabmayr01,Buchmann09a}.
 
Moreover,  based on the low $Q^2$ expansion of 
the neutron electric form factor, $G_{En} \simeq - \frac{1}{6} r_n^2 Q^2$, 
we can extrapolate the $Q^2$ dependence 
of the quadrupole form factors 
to~\cite{Grabmayr01,Buchmann04,Pascalutsa07a,Buchmann09a}
\ba
& &
G_E (Q^2)= \left(\frac{M}{M_\Delta} \right)^{3/2} 
\frac{M_\Delta^2 - M^2}{2 \sqrt{2}} \tilde G_{En} (Q^2)
\label{eqGE}\\
& &
G_C (Q^2)= \left(\frac{M}{M_\Delta} \right)^{1/2} 
\sqrt{2}
M M_\Delta  \tilde G_{En} (Q^2),
\label{eqGC}
\ea
where $\tilde G_{En}= G_{En}/Q^2$.
Hereafter we refer those results 
as large $N_c$ relations, 
since they  can be derived exclusively 
in the large $N_c$ limit~\cite{Pascalutsa07a}.

In this work we present improved large $N_c$ 
pion cloud parametrizations for the quadrupole 
form factors  in order to obtain a 
parametrization consistent simultaneously with Siegert's theorem 
and with the empirical data of the 
$\gamma^\ast N \to \Delta(1232)$ quadrupole form factors.

We conclude first that the relations (\ref{eqGE})--(\ref{eqGC})
implies that Siegert's theorem 
is violated by terms ${\cal R}_{pt} = {\cal O}(1/N_c^2)$, 
which may be a sizable error in the case $N_c=3$.
Since the relations (\ref{eqGE})-(\ref{eqGC}) are extrapolated from 
large $N_c$ they can have relative deviations 
of the order $1/N_c^2$.
We then use the constraints of Siegert's theorem
to modify the relation for $G_E$.
We obtain parametrizations 
for the quadrupole form factors 
that violate Siegert's theorem 
only by  terms ${\cal R}_{pt} = {\cal O}(1/N_c^4)$.
This result is thus compatible with Siegert's theorem 
apart from higher-order corrections in $1/N_c^2$.

We look also for additional contributions 
for the transition form factors $G_E$ and $G_C$, 
namely the contributions from the valence quarks 
from the nucleon and $\Delta(1232)$ systems.
As mentioned, those contributions are small 
in the context of quark models but 
combined with the parametrizations 
of the pion cloud contributions 
they can reduce the gap between theory and data.
An interesting propriety of the valence 
quark contributions for the electromagnetic 
form factors is that they vanish 
in the pseudo-threshold limit, as consequence 
of the orthogonality between the nucleon and $\Delta(1232)$ 
wave functions.
As a consequence, the test of Siegert's theorem 
condition ${\cal R}_{pt}=0$ needs to be tested 
only for the pion cloud contribution of the 
transitions form factors.

At the end, we combine valence and pion cloud contributions
using a model compatible with Siegert's theorem 
apart from higher-order corrections in $1/N_c^2$.
The results are then compared with the empirical 
data for $G_E$ and $G_C$, showing a fair description 
of the overall data.

\section{Pion cloud contributions}

We can test the quality of the relations 
(\ref{eqGE})--(\ref{eqGC}) comparing those 
functions with the data 
based on some parametrization for $G_{En}$.
To represent the electric form factor of the neutron,
we considers the Galster parametrization~\cite{Galster71} 
\ba
G_{En}(Q^2) = 
- \mu_n \frac{a \tau_N}{1 + d \tau_N}G_D,
\label{eqGEnX}
\ea 
where $\mu_n = -1.913$  
is the neutron magnetic moment,
$\tau_N= \frac{Q^2}{4M^2}$, $G_D= 1/(1 + Q^2/0.71)^2$ 
is the dipole factor,
and $a,d$ are two dimensionless parameters.

\begin{figure}[t]
\centerline{\mbox{
\includegraphics[width=2.6in]{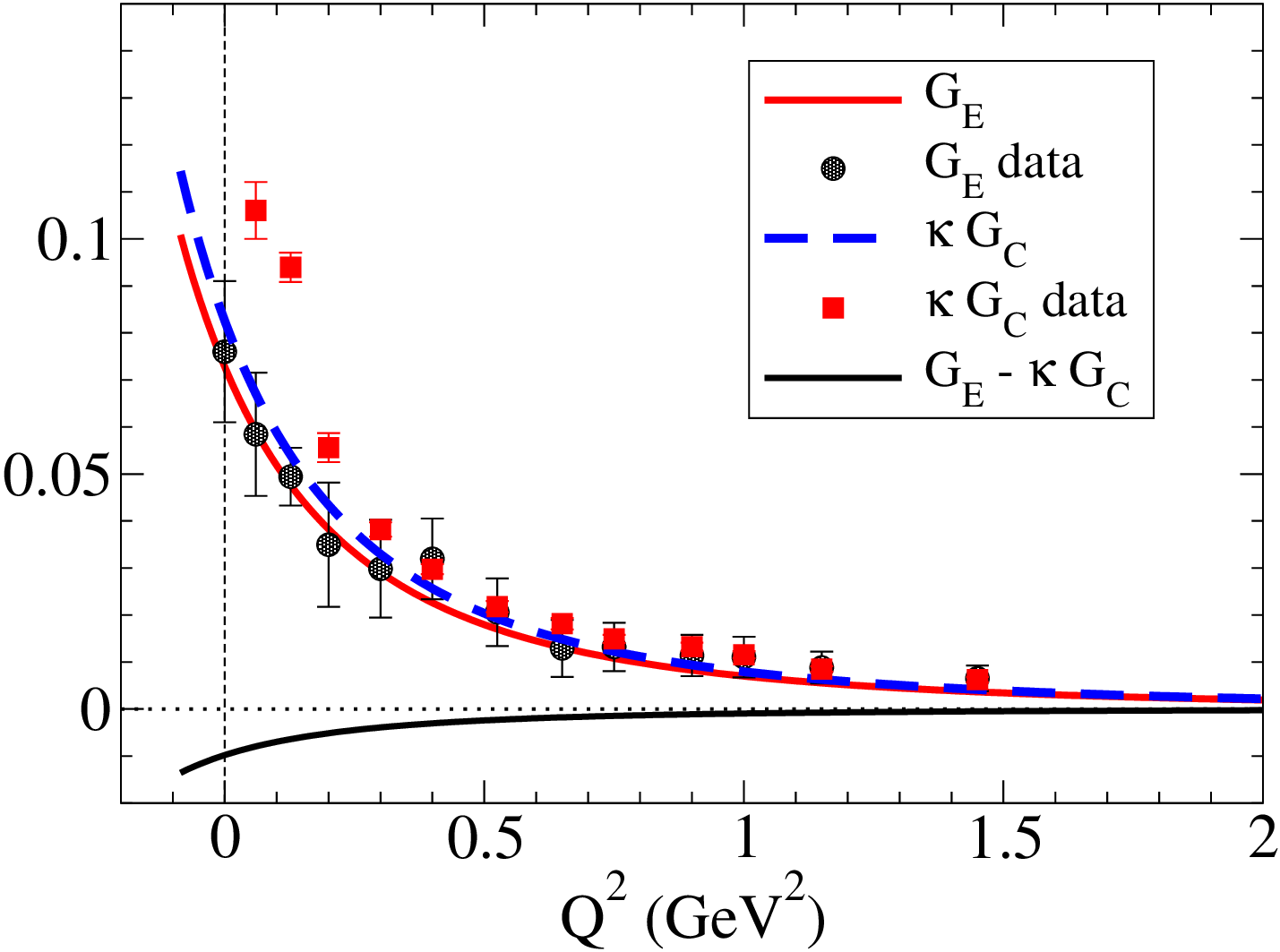}}}
\caption{\footnotesize
$\gamma^\ast N \to \Delta$ 
quadrupole form factors estimated by the  
pion cloud parametrization of Eqs.~(\ref{eqGE})-(\ref{eqGC}).
Data from Ref.~\cite{MokeevDatabase}.}
\label{figModel0}
\end{figure}

The quadrupole form factors  obtained
with the parameters $a=0.9$ and $d=2.8$~\cite{Buchmann09a}
are presented in Fig.~\ref{figModel0}.
For a better test of Siegert's theorem 
we multiply the function $G_C$ 
and the data for $G_C$ by $\kappa$.
The calculations are compared 
with the data from Mainz~\cite{Stave08},
MIT-Bates~\cite{MIT_data} and Jefferson Lab~\cite{Jlab_data} 
for finite square momentum transfer, $Q^2$,
and the world average from the 
Particle Data Group for $Q^2=0$~\cite{PDG}. 
The data are compiled in Ref.~\cite{MokeevDatabase}.

It is clear in Fig.~\ref{figModel0} that,  
the difference between the parametrizations for $G_E$ and $\kappa G_C$ 
is not zero in the pseudo-threshold limit, 
when $Q^2 \simeq -0.1$ GeV$^2$.
This result implies that Siegert's theorem 
is violated, because ${\cal R}_{pt} \ne 0$.
The explicit calculation
of the deviation 
using $G_{En}(Q_{pt}^2) \simeq - \frac{1}{6} r_n^2 Q_{pt}^2$, gives 
\ba
{\cal R}_{pt} \simeq -
\left( \frac{M}{M_\Delta} \right)^{3/2} \frac{r_n^2}{12 \sqrt{2}} Q_{pt}^2.
\ea
Since $Q_{pt}^2 = -(M_\Delta -M)^2$ 
and $M_\Delta - M = {\cal O}(1/N_c)$ 
we can conclude that ${\cal R}_{pt}= {\cal O}(1/N_c^2)$.
Although a result  
${\cal O}(1/N_c^2)$ may be seen 
as a small quantity, the numerical value is still sizable,
as we can see in the 
graph for ${\cal R}= G_E - \kappa G_C$
at the pseudo-threshold (${\cal R}_{pt}$).

\section{Valence quark contributions}

Before discussing how to improve 
the pion cloud parametrization of the 
quadrupole form factors $G_E$ and $G_C$, 
we may question if Siegert's theorem 
can in fact be verified for the valence quark sector.

We look then for the results obtained 
within valence quark models.
We consider, in particular, 
the covariant spectator quark model 
developed in Refs.~\cite{Nucleon,NDelta,NDeltaD,LatticeD,Omega} 
for the nucleon and $\Delta(1232)$ systems.
The basic assumptions of the model 
are that: (i) in the electromagnetic interaction 
the photon couples with the single quark 
(impulse approximation) while 
the other two quarks can be interpreted 
as an effective diquark, 
(ii) the quarks have their own 
internal structure (dressed by gluons and 
quark-antiquark states), and 
(iii) 
the radial quark-diquark wave functions
are calibrated in terms of  momentum range parameters 
that can be estimated by physical or lattice QCD data.

Concerning the nucleon and $\Delta(1232)$ systems 
the model is quite successful in the description 
of the data.
The parameters associated with the 
quark structure (quark electromagnetic form factors)
were first fixed by the nucleon elastic 
form factor data~\cite{Nucleon}.
After that the model was used to estimate 
the valence quark contribution 
for the $\gamma^\ast N \to \Delta$ 
magnetic dipole form factor~\cite{NDelta}.
The results are compatible simultaneously 
with estimates from dynamical reactions models~\cite{JDiaz07} 
and with lattice QCD simulations~\cite{NDelta,NDeltaD,Lattice,LatticeD}.
The model for the $\Delta(1232)$ was then extended with the 
inclusion of $D$ states in the
wave function~\cite{NDeltaD,LatticeD}.

We consider, in particular, the parametrization 
from  Ref.~\cite{LatticeD}.
In that work the $\Delta(1232)$ system is described 
by a combination a $S$-wave and two 
$D$-wave quark-diquark states.
The $D$ states can be labeled as $D3$ 
(quark total spin 3/2) and $D1$   
(quark total spin 1/2)~\cite{NDeltaD,NDelta}.
The nucleon system is represented just by a $S$ state~\cite{Nucleon}.
In the limit where the $D$-state mixtures vanish, 
we obtain $G_E \equiv 0$ and $G_C \equiv 0$.
The $D$-state mixtures and the 
$D$-radial wave functions are determined 
by a fit to the lattice QCD data from Ref.~\cite{Alexandrou08} 
and then extrapolated to the physical point.
The admixture of $D$ states is about 0.72\% for both states.
The extrapolation from lattice QCD for the physical 
regime is performed using a vector meson parametrization 
of the quark current (quark electromagnetic form factors).
More details can be found in Refs.~\cite{LatticeD,Omega,Lattice}.

\begin{figure}[t]
\centerline{\mbox{
\includegraphics[width=2.6in]{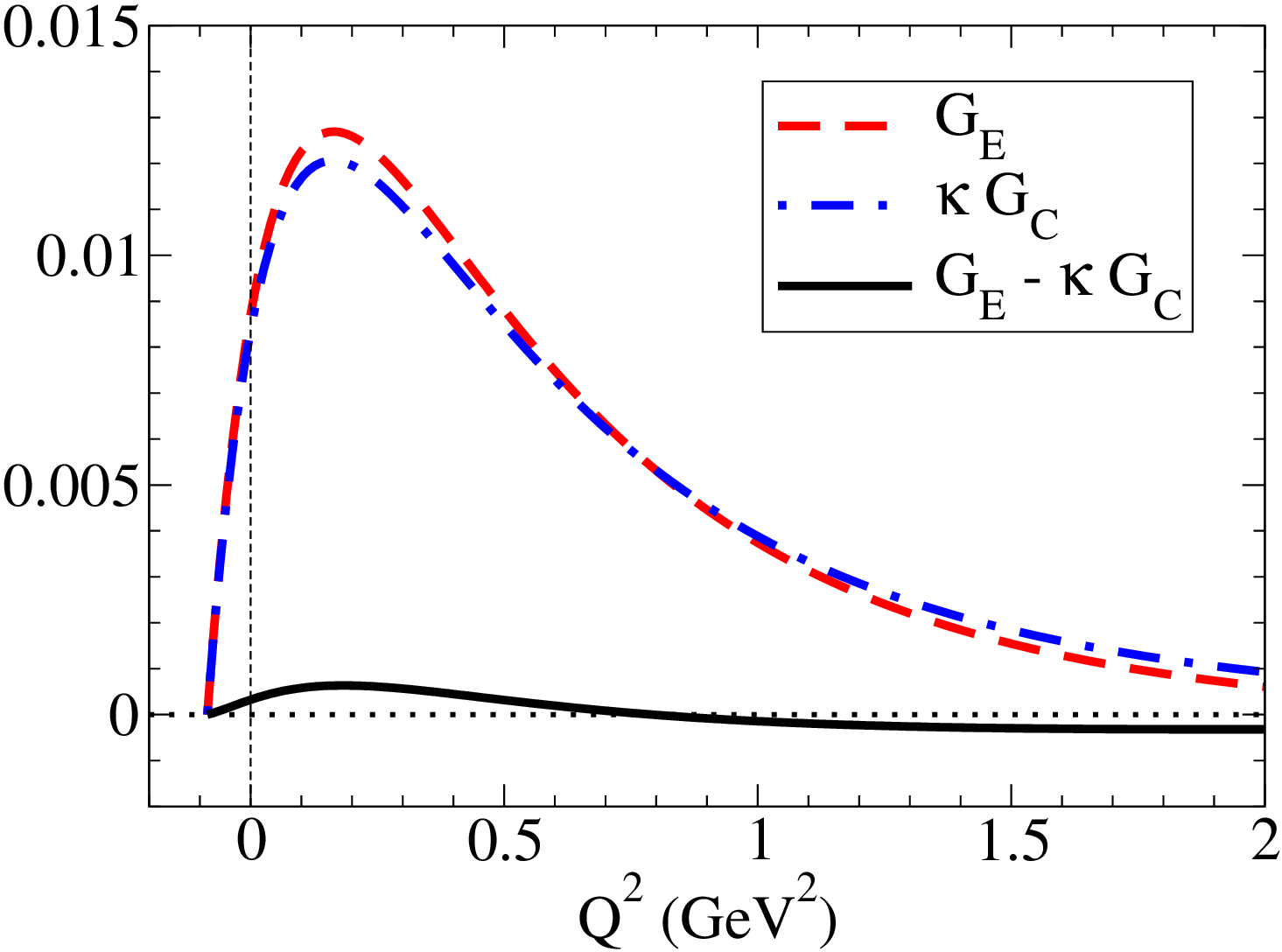}}}
\caption{\footnotesize
Valence quark contributions for the 
$\gamma^\ast N \to \Delta$ quadrupole form factors
estimated by the covariant spectator quark model~\cite{LatticeD}.}
\label{figVal}
\end{figure}

The valence quark contributions 
to the quadrupole form factors $G_E$ and $G_C$ from Ref.~\cite{LatticeD}
are presented in Fig.~\ref{figVal}.
In the figure, we again compare $G_E$ with $\kappa G_C$.
The more interesting aspect of the figure is that 
Siegert's theorem is exactly verified, 
as we can see from the result ${\cal R}= G_E - \kappa G_C=0$ 
at the pseudo-threshold.
This result is a consequence of $G_E=G_C=0$ at the same point.

In the covariant spectator quark model,
the quadrupole form factors $G_E$ and $G_C$ 
are calculated in terms of angular integrals of a function $b(k,q)$,
where $k$ and $q$ are, respectively, 
the diquark and the photon momenta.
In the pseudo-threshold limit ($|{\bf q}|=0$), 
the function  $b(k,q)$ reduces to 
the  spherical harmonic $Y_{20}(\hat {\bf k})$~\cite{NDeltaD}.  
The presence of the function $Y_{20}(\hat {\bf k})$ 
is then the consequence of the 
overlap between the $\Delta(1232)$ $D$ states and the nucleon $S$ state.
Since in the pseudo-threshold limit 
there is no dependence in the photon momentum $|{\bf q}|$, 
the angular integrals are reduced to the 
angular integration of $Y_{20}(\hat {\bf k})$,
which vanishes.
Consequently the form factors vanish too.
The result $G_E= G_C=0$ at the pseudo-threshold 
is then the corollary of the orthogonality 
between $S$ and $D$ states.

In Fig.~\ref{figVal}, we can notice  
the turning of the functions $G_E$ and $G_C$ near $Q^2 = 0.15$ GeV$^2$, 
just above the photon point $Q^2=0$
and a  soft convergence to zero at the pseudo-threshold. 
The reduction of the quadrupole form factors 
near $Q^2=0$ can  also be seen in the 
lattice QCD data~\cite{LatticeD,Alexandrou08}.
As we will see next, the presence of the maximum for $G_C$
near $Q^2=0.15$ GeV$^2$, instead at $Q^2=0$,  has implications 
in the values of $G_C$ at small $Q^2$.


Another interesting aspect concerning Fig.~\ref{figVal}
is the function  ${\cal R}= G_E - \kappa G_C$ 
for finite $Q^2$.
We can see that the function ${\cal R}$ is very small 
compared with $G_E$ or $\kappa G_C$.
We then concludes, that in the covariant 
spectator quark model, the results 
$G_E$ and $\kappa G_C$ are very similar.
It is possible that the relation $G_E \simeq \kappa G_C$
is also valid for other quark models.
We note, in particular, that  
the estimate of the quark core contribution 
used in the Sato-Lee model assumes 
$G_E = \kappa \frac{M_\Delta + M}{2 M_\Delta} G_C 
\simeq \kappa G_C$~\cite{JDiaz07}.

\section{Siegert's theorem}

We can now discuss if, as for the valence quark sector, 
the pion cloud parametrization 
is also  consistent with Siegert's theorem.

In the large $N_c$ limit the baryons 
are infinitely heavy and can be treated as static.
For this reason large $N_c$ is mostly 
used to calculate static proprieties 
of the baryon and transition between 
baryon states~\cite{Jenkins02,Dashen94}. 
In particular, the relations (\ref{eqGE})-(\ref{eqGC})
are extensions of large $N_c$ relations 
for $G_E$ and $G_C$  at $Q^2=0$ for finite $Q^2$~\cite{Pascalutsa07a}.
Those relations are not derived explicitly 
from  large $N_c$, but, since they 
derivation for $Q^2=0$ is based on large $N_c$,
one can infer that they are also limited by the accuracy 
from the large $N_c$ analysis, and can therefore 
be affected by relative corrections of the order 
$1/N_c^2$~\cite{Pascalutsa07a,Grabmayr01,Buchmann02,Buchmann09a,Jenkins02}.

Instead of using the $1/N_c$ expansion 
to calculate possible  $1/N_c^2$ relative corrections,
one can use Siegert's theorem 
to check if there are corrections that 
are consistent with the theorem.
One notes that the use of constraints 
external to the $SU(6)$ and large $N_c$ formalisms 
were used already in the calculation 
of coefficients associated with physical quantities as 
the charge radius, quadrupole moments and 
others~\cite{Buchmann02,Dillon89,Buchmann00a,Buchmann02b}.

In a first  attempt we checked if we can have 
an exact description of Siegert's theorem,
modifying the functions $G_E$ or $G_C$ 
keeping at the same time the results 
for $Q^2=0$, derived from large $N_c$.
Considering the replacement $G_E \to (1 + \alpha Q^2)G_E$,
we can preserve the result from large $N_c$ at $Q^2=0$, 
if we calculate $\alpha$ in order to obtain ${\cal R}_{pt}=0$ 
at the pseudo-threshold.
The solution for this condition is $\alpha= - \frac{1}{M_\Delta^2 -M^2}$.
We obtain the same effect if we replace $G_C \to G_C/(1+ \alpha Q^2)$.
The problem of the new form for $G_E$ is that it vanishes 
when $Q^2= M_\Delta^2 - M^2$,
in conflict  with the data.

We then try an approximated solution, replacing  
$G_E \to G_E/(1 + Q^2/(M_\Delta^2 - M^2))$, which 
induces no zeros for $Q^2 > 0$.
One obtain in this case
\ba
& &
G_E (Q^2)= \left(\frac{M}{M_\Delta} \right)^{3/2} 
\frac{M_\Delta^2 - M^2}{2 \sqrt{2}} 
\frac{\tilde G_{En} (Q^2)}{1 + \frac{Q^2}{M_\Delta^2-M^2}}. 
\label{eqGE2}
\ea
The previous relation differs from Eq.~(\ref{eqGE}) 
at the pseudo-threshold only by a term ${\cal O}(1/N_c^2)$.

With the new form for $G_E$,  one obtains 
\ba
{\cal R}_{pt} \simeq 
 \left( \frac{M}{M_\Delta} \right)^{3/2} \frac{M_\Delta-M}{2 M}
\frac{ r_n^2}{12 \sqrt{2}} Q_{pt}^2,
\ea
which is now a term ${\cal O}(1/N_c^4)$,
since $M_\Delta-M= {\cal O}(1/N_c)$ and $M = {\cal O}(N_c)$.

The expected falloff for large $Q^2$
of the pion cloud contributions for the 
form factors $G_E$ and $G_C$ 
given by Eqs.~(\ref{eqGC}) and (\ref{eqGE2})
are $G_E \propto 1/Q^8$ and $G_C \propto 1/Q^6$ 
respectively.
One recall however those
contributions are derived from the low $Q^2$ expansion 
of the neutron electric form factor 
and its application is in principle limited 
to intermediate values of $Q^2$~\cite{Pascalutsa07a}.  
The  high $Q^2$ region is expected 
to be dominated by the valence 
quark degrees of freedom~\cite{Carlson98} as discussed later.

To summarize, we use Siegert's theorem to 
find a correction for the form factor $G_E$
that minimizes the violation of Siegert's theorem.
The solution proposed, given by Eq.~(\ref{eqGE2}) is not exact,
but reduces the violation of  
Siegert's theorem to a term of the order $1/N_c^4$, 
The proposed function preserves the result for $G_E(0)$ 
obtained in the  large $N_c$ framework, 
and correspond to a relative correction 
of $1/N_c^2$ to the result of $G_E$ at the pseudo-threshold.

\section{Combination of pion cloud and valence quark contributions}

\begin{figure}[t]
\centerline{\mbox{
\includegraphics[width=2.6in]{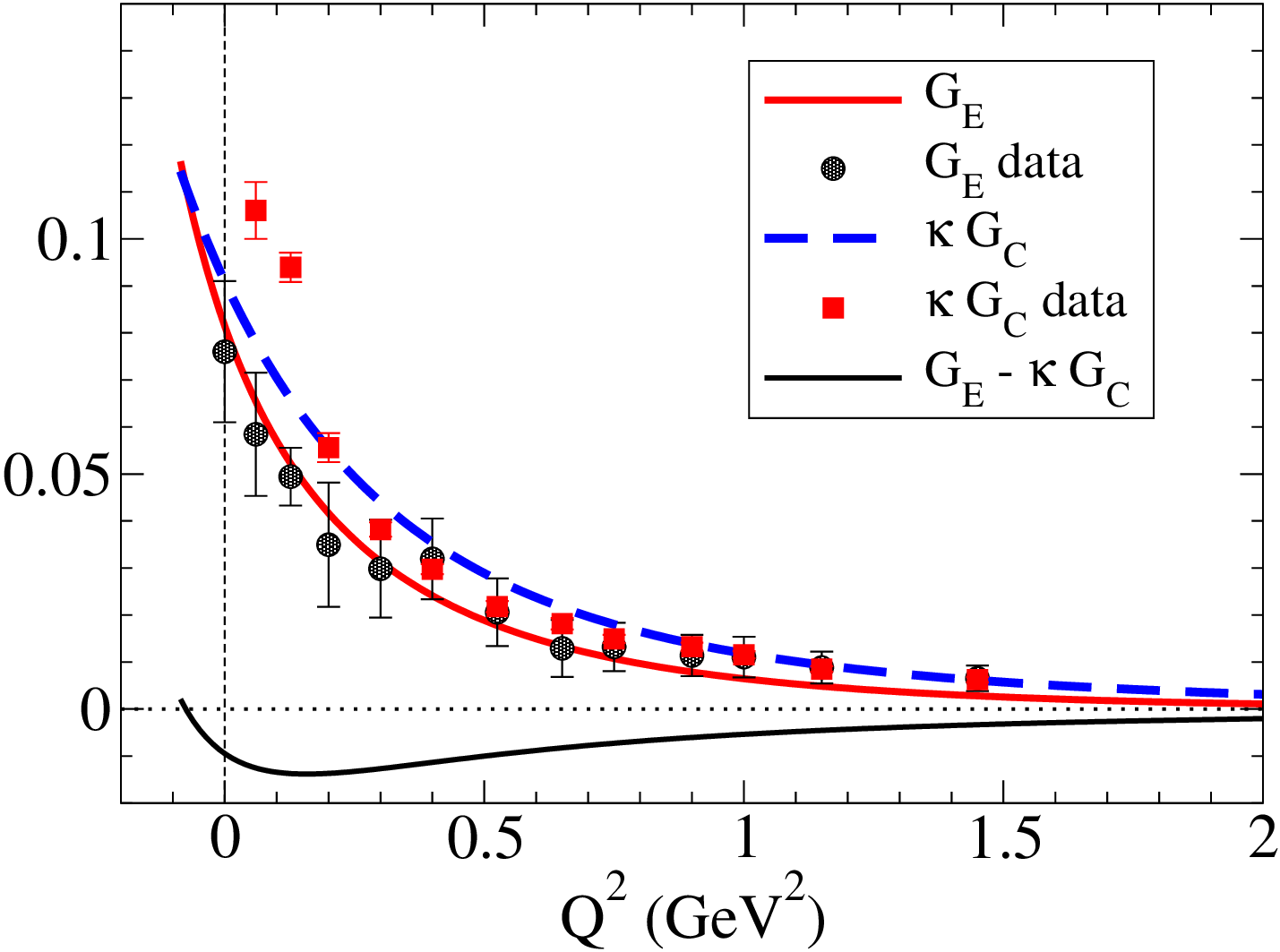}}}
\caption{\footnotesize
$\gamma^\ast N \to \Delta$ 
quadrupole form factors estimated 
as a combination of the 
valence quark and pion cloud contributions.
Data from Ref.~\cite{MokeevDatabase}.}
\label{figModel2}
\end{figure}

Since, as discussed, the pion cloud component 
represents only the leading order effect in $G_E$ and $G_C$,
we combine the new pion cloud parametrizations 
with the valence quark contributions of the covariant spectator 
quark model (consistent with Siegert's theorem).
The sum of the two contributions is presented 
in Fig.~\ref{figModel2}. 

From Fig.~\ref{figModel2}, we conclude that, 
apart from the results for $G_C$ below 0.2 GeV$^2$,
to be discussed later,
we obtain a very good description of the overall data.
This represents a considerable improvement 
comparative to the previous pion cloud parametrizations 
(see Fig.~\ref{figModel0}).
In addition, the form factors are now 
compatible with Siegert's theorem,
within an error 
of the order $1/N_c^4$.
The smallness of the error 
can be visualized in the figure 
if we look for ${\cal R}= G_E - \kappa G_C$
at the pseudo-threshold.

For future reference we call the attention 
for the fact that, the nonzero results 
for the form factors $G_E$ and $G_C$ are 
a direct consequence of the pion cloud contributions,
since as discussed, the valence quark contribution 
vanishes at the pseudo-threshold.

\begin{figure}[t]
\centerline{\mbox{\includegraphics[width=2.6in]{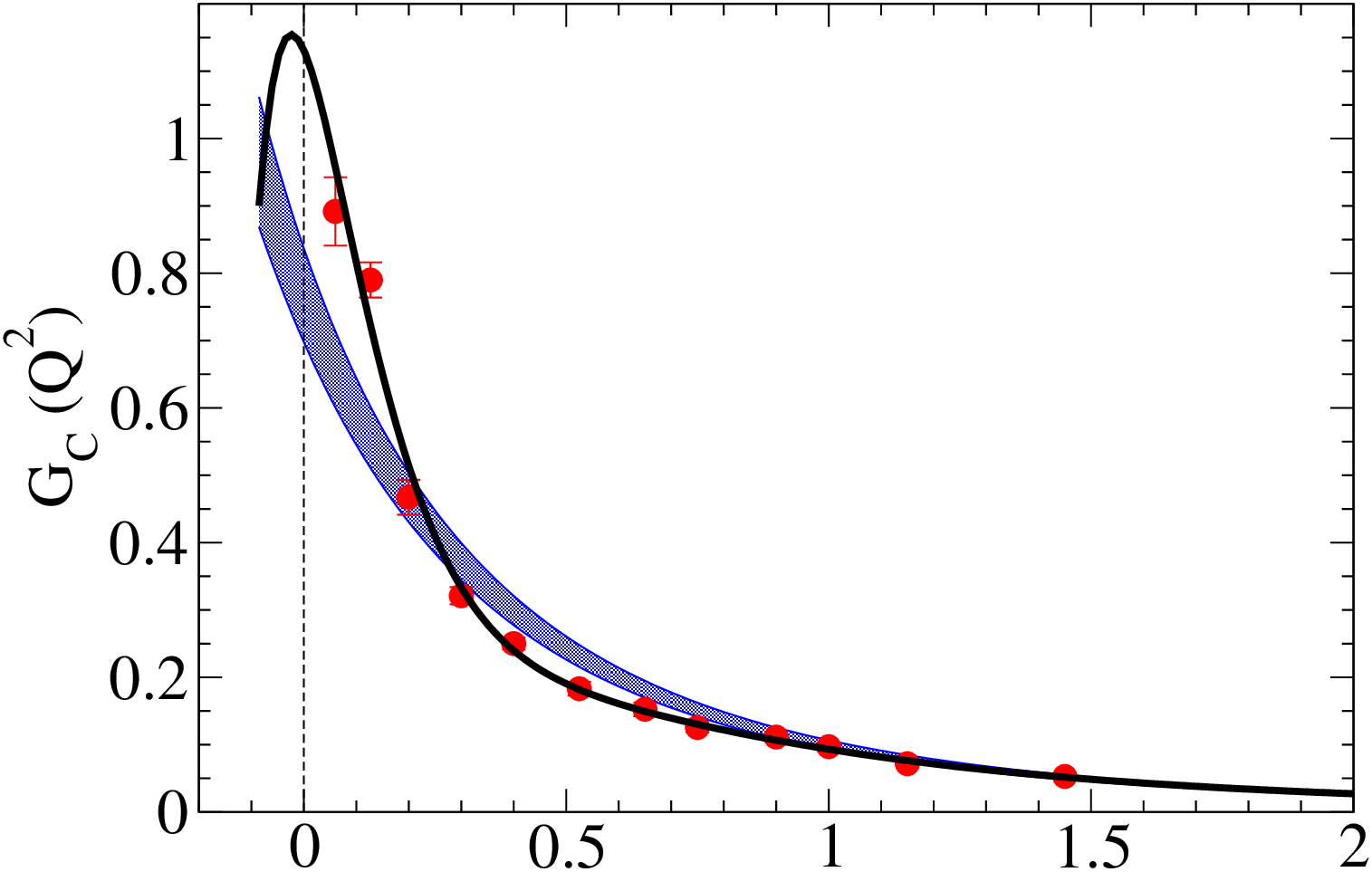}}}
\vspace{.9cm}
\centerline{\mbox{\includegraphics[width=2.6in]{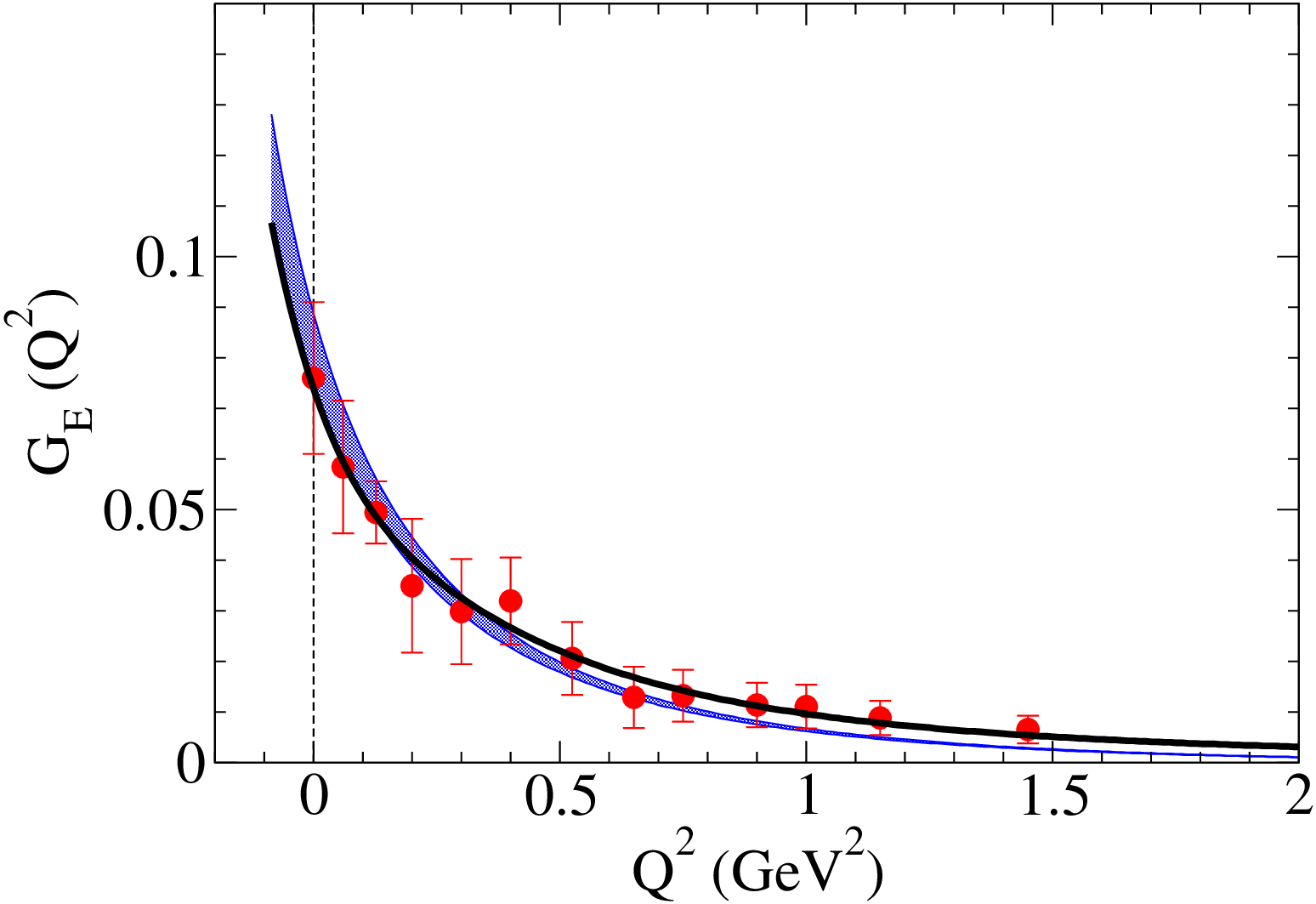}}}
\caption{\footnotesize
Results for the $\gamma^\ast N \to \Delta$ 
quadrupole form factors in comparison with the 
MAID-SG2 parametrization from Ref.~\cite{SiegertD}.
The bands indicate an estimate of 
the pion cloud error of 10\%, equivalent to 
the order of  $1/N_c^2$.
Data from Ref.~\cite{MokeevDatabase}.}
\label{figModel3}
\end{figure}

For a final discussion of the results, 
we need to take into account 
that the pion cloud contributions for $G_E$ and $G_C$ 
can have relative deviations of the order $1/N_c^2$
from Eqs.~(\ref{eqGC}) and (\ref{eqGE2}), as discussed previously.
To represent those deviations
we include a band of $\pm 10\%$ 
in the graphs for $G_E$ and $\kappa G_C$ 
to represent the possible relative deviation 
(term of order $1/N_c^2$) from 
the estimate of the pion cloud contribution.
Note that the  $\pm 10\%$ deviation 
is more relevant for the discussion of 
the spacelike region, where the data are available,
than near the pseudo-threshold. 
The final results for $G_E$ and $G_C$
including the variation band
are presented in Fig.~\ref{figModel3}.
In the figure we compare also the results 
with the MAID-SG2 parametrization  from Ref.~\cite{SiegertD},
in order to better visualize the difference 
between our model and the data at low $Q^2$.
The MAID-SG2 gives a high quality description of the data, 
and it is compatible with Siegert's theorem.

From the graph for $G_C$, 
we can conclude that, the gap between the 
present model and the data for $Q^2 < 0.2$ GeV$^2$ 
may not be explained by the 
pion cloud contribution, since the data are out of the band. 
We tested unsuccessfully if the 
quality of the description 
at low $Q^2$ could be improved 
considering a parametrization 
of the $G_{En}$ data more complex than 
the Galster parametrization (\ref{eqGEnX}).
Those results are an indication that the 
gap between model and data may be a consequence 
of the valence quark contributions.
Parametrizations of the quark core contributions
closer to a dipole form as in the Sato-Lee and DMT 
models from Ref.~\cite{JDiaz07,Kamalov99} are more appropriate
to describe the data measured at low $Q^2$.
Those parametrizations, however, differ  in shape, from 
the estimates presented in Fig.~\ref{figVal}, 
and are not compatible with the 
shape required by Siegert's theorem 
and the soft convergence to zero  at the pseudo-threshold.
A comparison between the results from 
the covariant spectator quark model and  
the parametrization from  Ref.~\cite{JDiaz07}
can be found in Ref.~\cite{LatticeD}.
As discussed in the context of 
the covariant spectator quark model, 
the convergence to zero at the pseudo-threshold
is a consequence of the orthogonality between 
the nucleon and the $\Delta(1232)$ states.

The shape  of the valence quark contributions 
in the region $Q^2=0$--0.2 GeV$^2$ 
can in principle be tested 
with the help of lattice QCD simulations.
With the advent of the lattice QCD simulations 
near the physical point, it is expected 
that in the near future the lattice QCD simulations 
approaches the physical point.
In those conditions, lattice simulations in quenched QCD
and  partially quenched QCD  approximations 
may be compared with our estimate of the valence quark contributions.
Also, the pion cloud contributions
can be estimated from the comparison 
between full QCD and quenched QCD.
An indication that the low $Q^2$ data for $G_C$ 
are reliable comes from the chiral effective-field theory,
which connects the lattice QCD data 
with large pion masses
with the physical regime~\cite{Pascalutsa05}.


Independent of the source of the 
discrepancy of $G_C$ for $Q^2 < 0.2$ GeV$^2$, 
the shape of the form factor $G_C$ has implications in the 
Coulomb quadrupole square radius.
Some authors argue that the low $Q^2$ behavior of $G_C$
is a consequence of the long extension of the 
pion cloud~\cite{Buchmann09a}.
Other works suggest instead, that,
as a consequence of Siegert's theorem,
$G_C$ becomes smoother near $Q^2=0$~\cite{SiegertD},
which according with the present analysis may 
be a consequence of the valence quark contributions,
since the  pion cloud contribution is sharper near $Q^2=0$.

In the range of the data presented $Q^2 < 2$ GeV$^2$ 
the pion cloud contributions are still sizable
(see Fig.~\ref{figModel0}).
As discussed the pion cloud parametrization 
are in principle valid for low $Q^2$ 
and may be modified for larger values of $Q^2$.
For $G_E$ the valence quark component 
falls as $1/Q^4$, as expected from a quark model~\cite{NDeltaD,LatticeD},
and dominates over the pion cloud component ($1/Q^6$).
As for $G_C$ the valence quark component 
falls as $1/Q^6$, as the pion cloud component.
The final falloff is then 
$G_E \propto 1/Q^4$ and 
$G_C \propto 1/Q^6$, respectively, 
consistent with pQCD~\cite{Carlson98}.
We conclude then that, for very large $Q^2$, the present model 
is compatible with the expected pQCD falloff.
Recall that the pQCD falloff is 
the consequence of the dominance of the
valence quark contributions~\cite{Carlson98}.
Another pQCD prediction is that, 
$G_E \simeq - G_M$, for very large $Q^2$~\cite{Carlson98,Carlson86}.
Experimentally we are nowadays far away 
from this result~\cite{NSTAR,Carlson98}.

Overall, our calculations support the idea 
that the physics associated with the 
$\gamma^\ast N \to \Delta(1232)$ transition 
can be described by adding pionic degrees of freedom 
to the quark models~\cite{NSTAR,QpionCloud1,QpionCloud2}.
Dynamical reaction models such as the Sato-Lee model~\cite{JDiaz07}
and the DMT model~\cite{Drechsel2007,Kamalov99},
which calculate the pion cloud dynamically,
are also in qualitative agreement  
with the data.
In those models the bare core contributions are estimated 
phenomenologically as  already discussed
for the case of Ref.~\cite{JDiaz07}.

\section{Summary and conclusions}

In conclusion, we have proposed 
a new  pion cloud parametrization 
for the $\gamma^\ast N  \to \Delta(1232)$ 
electric quadrupole form factor.
The new form for $G_E$ is consistent 
with Siegert's theorem, 
$G_E = \kappa G_C$, at the pseudo-threshold,
within an error of $1/N_c^4$.

We have also discussed  the implications of 
Siegert's theorem to the bare core 
contribution of the quadrupole form factors.
Contrary to the pion cloud contributions, 
the valence quark contributions vanish 
at the pseudo-threshold.

Combining the new parametrizations of 
pion cloud contributions with the 
valence quark contributions for the same form factors, 
we have obtained a very accurate description of the 
quadrupole form factors data,
apart from a small underestimation of 
the $G_C$ data in the region $Q^2=0$--0.15 GeV$^2$.

Future developments in lattice QCD 
may help to clarify if the the gap between theory and data
at low $Q^2$ 
is a consequence of the underestimation 
of the valence quark component or 
of the pion cloud component.

\begin{acknowledgments}
The author thanks Jo\~ao Pacheco B.~C.~de Melo 
and Kazuo Tsushima for the hospitality 
at Universidade Cruzeiro do Sul, where this work started,
and Pulak Giri for useful suggestions.
This work is supported by the Brazilian Ministry of Science,
Technology and Innovation 
and by the project 
``Science without Borders'' from the Conselho Nacional
de Desenvolvimento Cient\'{i}fico e Tecnol\'ogico (CNPq) 400826/2014-3.
\end{acknowledgments}

\end{document}